\documentclass[12pt]{article}
\usepackage{graphicx}

\newcommand{\Det}{{\rm Det}}
\newcommand{\Tr}{{\rm Tr}}

\newcommand{\bra}[1]{\left\langle{#1}\right|}
\newcommand{\ket}[1]{\left|{#1}\right\rangle}
\newcommand{\ave}[1]{\left\langle{#1}\right\rangle}

\newcommand{\be}{\begin{equation}}
\newcommand{\ee}{\end{equation}}
\newcommand{\bea}{\begin{eqnarray}}
\newcommand{\eea}{\end{eqnarray}}
\newcommand{\ba}{\begin{array}{l}}
\newcommand{\ea}{\end{array}}
\newcommand{\half}{\frac{1}{2}}
\newcommand{\gf}{\gamma_5}

\begin{document}

\begin{center}

LOW ENERGY CONSTANTS OF CHIRAL PERTURBATION THEORY \\
FROM THE INSTANTON VACUUM\\
M.~Musakhanov\\
National University of Uzbekistan
\end{center}
\begin{abstract}
In the framework of the instanton vacuum model we make expansion over the current mass $m$ and number of colors $N_c$ and evaluate ${\cal O}(1/N_c,\,m,\,m/N_c,\,m \,\ln m/N_c)$ corrections to the  dynamical quark mass $M$, the quark condensate $\langle\bar qq\rangle$, the pion mass $M_\pi$ 
and decay constant $F_\pi$. We found the $SU(2)$ $\chi$PT low-energy constants $\bar l_3, \bar l_4$ in a good correspondence with the phenomenology and lattice calculations.
\end{abstract}
{\bf Introduction.}
 In QCD with massless $u,d$ quarks  the $SU(2)\times SU(2)$ chiral symmetry
is spontaneously broken and leads to appearance of the massless Goldstone particles -- pions.
In reality, $u,d$ quark current masses are non-zero but small on the hadronic scale $\sim 1\,GeV.$

The Chiral Perturbation Theory ($\chi$PT) was proposed in~\cite{Gasser:1983yg}  for parameterization of the QCD hadronic correlators at low-energy region, where the expansion parameters are  light quark current masses $m$ and pion momenta $p$. The basic tool is the phenomenological effective lagrangian, which has a form of the infinite series in these parameters. Naturally, the low-energy constants (LEC) of the series expansion are not fixed. Up to now they were extracted only from the experimental data. Recent progress in lattice calculations provide us with the estimates of LEC. The main problem of lattice evaluations is the still-large pion masses $M_\pi$ available on the finite size lattices. 

QCD instanton vacuum model, often referred  as the instanton liquid model, provides a very natural nonperturbative explanation of the S$\chi$SB (see the reviews ~\cite{Schafer:1996wv,Diakonov:2002fq}). It provides a consistent framework for description of the pions and thus may be used for evaluation of the LEC.
  Quasiclassical considerations show that it is energetically favourable to have lumps of strong gluon fields (instantons) spread  over 4-dimensional Euclidian space. Such fields do strongly modify the quark propagation due to the t'Hooft type quark-quark interactions in the background of the instanton vacuum field. This background is assumed as a superposition of $N_+$ instantons and $N_-$ antiinstantons 
$
 A_\mu (x) = \sum^{N_+}_I A^I_{\mu}(\xi_I, x)
 +\sum^{N_-}_{A} A^{A}_{\mu}(\xi_{A}, x),
 $
 where $\xi=(\rho,z,U)$ are the (anti)instanton collective coordinates -- size, position and color orientation. The most essential for the low-energy processes are the would-be quark zero modes, which result in a very strong attraction in the channels with quantum numbers of vacuum, appearance of the quark condensate
 and generation of the dynamical quark mass.  
 The main parameters of the model are the average inter-instanton distance $R$ 
 and the average instanton size $\rho$. The  estimates of these quantities are
$\rho\simeq   0.33 \, fm, \, R\simeq 1\, fm$ (phenomenological), 
$\rho\simeq 0.35\, fm,\, R \simeq 0.95\, fm$ (variational)~\cite{Schafer:1996wv,Diakonov:2002fq},
$\rho\simeq 0.36\,fm,\,   R\simeq 0.89\, fm$ (lattice)~\cite{Chu:vi}-\cite{Bowman:2004xi} 
 and have $\sim 10-15\%$ uncertainty.
Recent computer investigations~\cite{Negele0605256} of a current mass dependence of QCD observables within instanton liquid model show that the best correspondence with lattice QCD data is obtained for 
$\rho\simeq  0.32\, fm,\,\,\,\, R\simeq  0.76\, fm.$
While in the real world the number of colors is $N_c=3$, since the pioneering work of t'Hooft it is \textit{assumed} that one can consider  $N_c$-counting  as a useful tool, \textit{i.e.} take the limit $N_c\to\infty$ and neglect the $1/N_c$-corrections. In the instanton vacuum model $N_c$-counting is naturally incorporated.
The phenomenological set is popular since in the leading order (LO) it reproduces reasonable values for most of the physical quantities. This leads to rather consistent description of pions and nucleons in the chiral limit.

The main purpose of this work is evaluation of ${\cal O}(1/N_c,\,m,\,m/N_c,\,m/N_c\,\ln m)$ non-leading-order (NLO) corrections to different physical observables, which provide LECs. So, we are dealing with double expansion over $m$ and over $1/N_c.$
There are several sources of such NLO corrections:\\
1. At pure gluonic sector of the instanton vacuum model the width of the instanton size distribution is ${\cal O}(1/N_c)$. The account of the finite width leads to rather small corrections
. In the following we will check the accuracy of $\delta$-function type of the instanton size distribution by direct evaluation of the finite width corrections.
\\
2. The back-reaction of the light quark determinants to the instanton vacuum properties is formally controlled by $N_f/N_c$-factor
. It does not sizably change the distribution over $N_++N_-$ but radically change the distribution over $N_+ - N_-$. Any $m_f=0$ leads to $\delta$-function type of the distribution 
. In the following we take $N_+ = N_-$.
\\ 
3. {There are the quark-quark tensor interaction terms which are $1/N_c$-suppressed and thus are absent in the $N_c$LO effective action. These terms correspond to nonplanar diagrams in old-fashioned diagrammatic technique.}
\\ 
4. The contribution of meson quantum fluctuations (meson loops) has to be taken into account.
\\
 First we study the role of the meson loops which give the dominant contribution. At the end we also estimate the contributions of finite width of instanton size distribution, and  above-mentioned tensor interaction term.
\\
We consider parameters $\rho, R$ as free within their $\sim 15\%$ uncertainty and fix them from the requirement $F_{\pi,{m=0}}=88 MeV, \langle\bar qq\rangle_{m=0}=(255 MeV)^3$ with account of $N_c$NLO corrections, as it is requested by $\chi$PT. We found the values
 $\rho=0.350 fm,\,\,R=0.856 fm$
 in agreement with the above-given estimates.
Note that though the evaluation of the meson loop corrections in the instanton vacuum model is similar to the earlier meson loop evaluations \cite{Nikolov:1996jj}-\cite{Oertel:2000cw}
in the NJL model, there are a few differences:\\
1. As it has been already  mentioned, the meson loop corrections are not the only sources of $1/N_c$-corrections in the instanton model.
\\
2. Due to nonlocal form-factors there is no need to introduce independent fermion and boson cutoffs $\Lambda_f, \Lambda_b$. The natural cutoff scale for all the loops (including meson loops) is the inverse instanton size $\rho^{-1}$.
\\ 
3. The quark coupling constant is defined through the saddle-point equation in the instanton model whereas it is a fixed external parameter in NJL.

The basic object we study are the correlators. 
The simplest correlators are: $\ave{qq(m)}=-F^2 B+{\cal O}(m)$,    
$\int d^4 x\, e^{-iq\cdot x}\ave{j_\mu^{a,5}(x)j_\nu^{b,5}(0)}=
F_\pi^2\delta^{ab}\left(g_{\mu\nu}-\frac{q_\mu q_\nu}{q^2+M_\pi^2}\right)+{\cal O}(q^2)$, 
$F_\pi= F+{\cal O}(m)$, $M_\pi^2= m_\pi^2+{\cal O}(m^2),$ where 
$m_\pi^2 = 2 B\, m$.                             
The constants $F, B$ define pion decay constant and quark condensate in the chiral limit, which are the $\chi$LO LECs.          
 $\chi$PT provide proper way of the parameterization of the correlators (observables) 
 in low energy region  by means of $\chi$LO LECs $F,B$ and $\chi$NLO LECs $ l_i$, defined in ~\cite{Gasser:1983yg}. 
 Important fact that  \textit{bare} constants $l_i$ are renormalized by pion loops which lead to 
 $l_i\rightarrow \bar l_i.$ The correlators and observables of pion physics should be expressed in terms of $\bar l_i$. 
                                                                                          
There is a number of running experiments dedicated to the low-energy pion physics:
$\pi^+\pi^-, \pi K$ atoms @DIRAC@CERN, $K\rightarrow\pi\pi e\nu$ @BNL E865, $K^\pm\rightarrow\pi^\pm\pi^+\pi^-$ @NA48/2, pion electromagnetic polarizabilities @Mainz Microtron  
MAMI, etc. and
QCD lattice evaluations of LECs - the collaborations MILC, ETM, JLQCD, RBC/UKQCD, PACS-CS~\cite{MILC}-\cite{PACS-CS}.

We assume that the promising method of the calculation of LECs is the application of instanton vacuum model.

{\bf Light quarks in instanton vacuum \cite{Musakhanov1999}-\cite{Goeke:2007}.}
Zero-mode approximation for the quark propagator in a single instanton field --
$                                                                                    
 S_i(x,y)\approx \frac{\ket{\Phi_{0i}}\bra{\Phi_{0i}}}{im}+ \frac{1}{i \hat \partial},\,\,\,\,
 ((i\hat\partial + g \hat A_{i} )\Phi_{0i}=0),                                          
$
is working well at $m\Rightarrow 0$ \cite{Diakonov:2002fq} but wrong beyond the chiral limit.            
Our extension \cite{Musakhanov1999}-\cite{Goeke:2007} of zero-mode approximation beyond the chiral limit:                                                
\bea                                                                                                          S_i=S_{0} - S_{0}\hat p \frac{|\Phi_{0i}\rangle\langle\Phi_{0i}|}{\langle\Phi_{0i}|\hat p S_{0} \hat p       
|\Phi_{0i}\rangle} \hat p S_{0},\,\,\, 
S_0=\frac{1}{\hat p +im},                                                                               
\,\,\,                                                                                                    S_i|\Phi_{0i}\rangle = \frac{1}{im}|\Phi_{0i}\rangle,\,\,\, \langle\Phi_{0i}|S_i                             
 =\langle\Phi_{0i}|\frac{1}{im}.                                                                             
\eea 
Then, quark propagator in instanton media and in the presence of the external fields $\hat V=s+p\gamma_5+\hat v+\hat a\gamma_5$ become:           
\begin{eqnarray}                                                                                             
&&\tilde S -\tilde S_{0} = -\tilde S_{0}\sum_{i,j}\hat p |\phi_{0i}\rangle                          
\left\langle\phi_{0i}\left|\left(\frac{1}{\hat p \tilde S_{0}\hat p}\right)\right|\phi_{0j}\right\rangle     
\langle\phi_{0j}|\hat p \tilde S_{0}\\                                                                       
&&\nonumber|\phi_0\rangle=\frac{1}{\hat p}L \hat p |\Phi_0\rangle,\,\,\,                                     
\tilde S_{0}=\frac{1}{\hat p +\hat V+im },\,\,\,
\nonumber L_i(x,z_i)={\rm P} \exp\left(i\int_{z_i}^x dy_\mu( v_\mu(y)+a_\mu(y)\gamma_5)\right)             
\end{eqnarray}                                                                                               
 From this one the low-frequencies part of quark determinant is:
 \bea
 \ln \tilde\Det_{low}=\Tr\int dm\, \tilde S(m) =\ln \det \langle\phi_{0,i}|\hat p \tilde S_0^{fg}  
 \hat  p|\phi_{0,j}\rangle,                                                                                      
 \eea
  Averaging of $\tilde\Det_{low}$ over instantons by means of fermionization leads to the  partition function $Z_N$ in terms of constituent quarks $\psi$ (in the following $N_f=2).$ Then,
\bea  
&&Z_N=\int d\lambda_+d\lambda_-D\bar\psi D\psi e^{-S}  
,\,\,S=\sum_\pm\left(N_\pm \ln\frac{K}{\lambda_\pm}-N_\pm+
\psi^\dagger (i\hat \partial+\hat V+im)\psi + \lambda _\pm Y_2^\pm \right)
\\
&&\nonumber 
  Y_2^\pm=\int d\rho D(\rho)\left(\alpha^2 \det_f J^\pm+\beta^2 \det_f J^\pm_{\mu\nu}\right)
 ,\,\,\,\frac{\beta^2}{\alpha^2}:= \frac{1}{8N_c}\frac{2 N_c}{2 N_c-1}=\frac{1}{8N_c-4}={\cal O}\left(\frac{1}{N_c}\right)\\ 
  &&\nonumber J^\pm_{fg}=\psi^{\dagger}_f\bar L \frac{1\pm \gf}{2}L\psi_g,\;
J^\pm_{\mu\nu}=\psi^{\dagger}_f\bar L \sigma_{\mu\nu}\frac{1\pm \gf}{2}L\psi_g.
\eea
Here the dynamical quark-quark interaction coupling $\lambda_\pm$ 
is due to the exponentiation in $Z_N$ by using of Stirling-like formula.
Further step -- the bosonization in terms of mesons is obvious procedure for $N_f=2$ case. The 
integration over fermions provide  the action:
\begin{eqnarray}
&& S=-N\ln\lambda+2\sum_i\left(\Phi_i^2+\half\Phi_{i,\mu\nu}^2\right)
\\\nonumber
&&- Tr\log \left[\hat p+\hat V+im + 
i \lambda^{0.5}\bar L F(p)\left(\alpha \Phi_i\Gamma_i+\half\beta \Phi_{i,\mu\nu}\sigma_{\mu\nu}\Gamma_i\right) F(p) L^{-1}\right]
\end{eqnarray}
The mesons $\Phi_i,\Phi_{i,\mu\nu}$ are chiral doublets: 
$(\sigma, \vec \phi)$, $(\vec \sigma,\eta)$ and $(\sigma_{\mu,\nu},\vec \phi_{\mu\nu})$.

{\bf Dynamical quark mass \cite{Musakhanov2001}, \cite{Kim:2005jc}, \cite{Goeke:2007bj}.}
For evaluation of the partition function $Z_N$, it is  used  the effective action \cite{Coleman:1973jx,Jackiw74} $\Gamma_{eff}[m,\lambda,\sigma]$, defined as:
\begin{eqnarray}
Z_N[m]=\int d\lambda Z_N[m,\lambda]=\int d\lambda\exp(-\Gamma_{eff}[m,\lambda,\sigma])
\label{Veff}
\end{eqnarray} 
where the vacuum field  $\sigma$ and the coupling $\lambda$ are the solutions of the Eqs.
\begin{eqnarray}
\frac{\partial \Gamma_{eff}[m,\lambda,\sigma]}{\partial\sigma}=0,\,\,\,\,\,
\frac{\partial \Gamma_{eff}[m,\lambda,\sigma]}{\partial\lambda}=0.
\label{vacuum}
\end{eqnarray}
In the $N_c$LO $\Gamma_{eff}[m,\lambda,\sigma]=S[m,\lambda,\sigma].$ Meson fluctuations provide NLO
term given by
\begin{eqnarray}
\label{Vmes}
\Gamma_{eff}^{mes}[m,\lambda,\sigma]=\frac{1}{2}\Tr\ln\left( 4\delta_{ij}-\frac{1}{\sigma^2}
\Tr \frac{M(p)}{\hat p+ i \mu(p)} \Gamma_i \frac{M(p)}{\hat p+ i \mu(p) }\Gamma_j\right),
\end{eqnarray}
where $\mu(p)=m+M(p)$ and we introduced the dynamical quark mass $M(p)=M F^2(p)$; $M=\frac{(2\pi\rho)^2\lambda^{0.5}}{2g}\sigma.$
The dynamical quark mass $M(p)$ and the coupling $\lambda$ are the solutions of the vacuum and saddle-point Eqs.. The numerical solution of the Eqs. is
\bea
M(m)=0.36-2.36\,m -\frac{m}{N_c}(0.808+4.197 \ln m)
\eea
Here and in the following $M$ and $m$ are given in $GeV.$
 
 {\bf Quark condensate \cite{Musakhanov2001}, \cite{Kim:2005jc}, \cite{Goeke:2007bj}.}
The quark condensate $\langle\bar qq\rangle$  characterizes the S$\chi$SB  and is given by Eq. $\langle\bar qq\rangle =\frac{\partial \ln Z_N}{\partial m}$
Numerical evaluation  gives 
 \begin{eqnarray}
 -\langle\bar qq\rangle (m) =\left(\left( 0.00497 - 0.0343m \right) N_c+
 +\left(0.00168 - 0.0494m
 - 0.0580m\ln m\right)\right)[GeV^3]
 \end{eqnarray}

{\bf Quarks in  external axial-vector field and pion properties \cite{Goeke:2007bj}.}
External axial-vector isovector field $a_\mu=a_{\mu}^{i}\tau_i /2$ generate nonzero vacuum pion field 
$\vec u$ and we have an additional vacuum equation:
\begin{eqnarray}
\frac{\partial\Gamma_{eff}[m,\lambda, \vec u, \vec a_\mu]}{\partial\vec u}=0.
\label{vacuumSigmau}
\end{eqnarray}
We may easily get that the shifts of $\sigma, \lambda$ contribute only to ${\cal O} (a^4)$-terms and thus may be safely omitted.
The total vacuum meson fields are represented as $\Phi_{vac}=\sigma\, U,\,\,\, U=u_0+i\vec\tau\vec u,\,\, U^\dagger U=UU^\dagger=1$. In the NLO one has to take into account the fluctuations of the meson fields 
$\Phi= \Phi_{vac} + \Phi'.$ Now $\Gamma_{eff}[m,\lambda, \vec u, \vec a_\mu]=S[m,\lambda, \vec u, \vec a_\mu]+ \Gamma^{mes}_{eff}[m,\lambda, \vec u, \vec a_\mu],$ where 
\begin{eqnarray}
\Gamma^{mes}_{eff}[m,\lambda, \vec u, \vec a_\mu]=
\frac{1}{2}\Tr\ln \frac{\delta^2  S[m,\lambda,\sigma,\vec u,\vec a_\mu,\Phi']}{\delta\Phi'_i
\delta\Phi'_j}|_{\Phi' =0}
=
\Gamma^{mes}_{eff}[m,\lambda]+\Delta\Gamma^{mes}_{eff}[m,\lambda, \vec u, \vec a_\mu]
\end{eqnarray}
The first term was calculated before and we have to calculate now second one.
Collecting the terms $a_\mu a_\nu$, $a_\mu \partial_\nu u_i$ and $\partial_\nu u_i\partial_\mu u_j$, we show that in agreement with chiral symmetry 
$
\Gamma_{eff}=
F_{aa}^2 \vec a_\mu^2+F_{uu}^2  (\partial_\mu \vec u)^2+2 F_{au}^2 \vec a_\mu\partial_\mu \vec u + F_{uu}^2  M_\pi^2 \vec u^2
+{\cal O}(a^3,u^3, m^2),
$
where the constants $F_{ij}$ differ only beyond chiral limit:
$
F_{aa}^2 -F_{uu}^2=2\left(F_{au}^2 -F_{uu}^2\right)
\sim m.
$
Then, 
one can get that the two-point axial-isovector currents correlator has a form:
\begin{eqnarray}
\label{aa:structure}
\int d^4 x e^{-iq\cdot x}\langle j^{A,i}_\mu(x)j^{A,j}_\nu(0)\rangle=
=\delta_{ij}F_\pi^2\left(\delta_{\mu\nu}-\frac{q_\mu q_\nu}{q^2+M_\pi^2}\right)+{\cal O} (q^2)
\end{eqnarray}
We see that  $M_\pi$ has a meaning of pion mass and $F_\pi$ -- pion decay constant.

{\bf Pion decay constant $F_\pi$ and mass $M_\pi$ from $\Gamma_{eff}$ \cite{Goeke:2007bj}.}
Finally numerical calculations lead to
\begin{eqnarray}
\label{Res:Fpi}
F_\pi^2=N_c\left(\left(2.85 -\frac{0.869}{N_c}\right)-\left(3.51+\frac{0.815}{N_c}\right)m
-\frac{44.25}{N_c}\,m\,\ln m +{\cal O} (m^2)\right)\cdot 10^{-3}\; [GeV^2]
\end{eqnarray} 
and
\begin{eqnarray}
\label{Res:Mpi}
M_\pi^2=m\left(\left(3.49+\frac{1.63}{N_c}\right)+m\left(15.5+\frac{18.25}{N_c}
+\frac{13.5577}{N_c} \ln m \right)+{\cal O}(m^2)\right)
\end{eqnarray}
Chiral log theorems~\cite{Gasser:1983yg} provided the test of all of the numerical calculations above. We found that all nonanalytical $m\ln m$ terms in $\langle\bar qq\rangle$ , $F_\pi$ and $M_\pi$ very well correspond them.

{\bf Instanton finite width and tensor terms contributions to $F_\pi$, $M_\pi$~\cite{Goeke:2007bj}.}
The main effect of the finite width is the change of the $p$-dependence of dynamical quark mass $M(p).$
For the estimations we take $\delta\rho^2=\ave{\rho^2}-\ave{\rho}^2\approx\frac{0.5599\, GeV^{-2}}{N_c}$ 
\cite{Diakonov:2002fq}
which leads to the 
$\approx 5\%$ for $F_\pi^2$ and $\approx 2.6\%$ for $\ave{\bar qq}$ corrections. They are negligible one.
Tensor terms do not affect $F_\pi$ and $M_\pi$, since they lead to the $O(q^2)$-correction to the axial currents correlator. 

{\bf LEC $\bar l_3$ and $\bar l_4$} provide $\chi$NLO terms in $F_\pi(m)$ and $M_\pi(m)$ as
\begin{eqnarray}
 M_\pi^2=m_\pi^2\left(1-\frac{m_\pi^2}{32\pi^2 F^2}\bar l_3+{\cal O}(m_\pi^4)\right),\,\,\,\,
 F_\pi^2=F^2\left(1+\frac{m_\pi^2}{8\pi^2 F^2}\bar l_4+{\cal O}(m_\pi^4)\right),
 \end{eqnarray}
and can be extracted from our previous estimations. They are compared with lattice calculations ~\cite{MILC}-\cite{PACS-CS} and phenomenological estimations~\cite{Gasser:1983yg} at the following
Figs.\footnote{We added our results to the  phenomenological and lattice data Figs. from~\cite{Leutwyler2008}.}, where left panel represent $\bar l_3$ and right one represent $\bar l_4$.

\centerline{\includegraphics[scale=0.3,angle=-90]{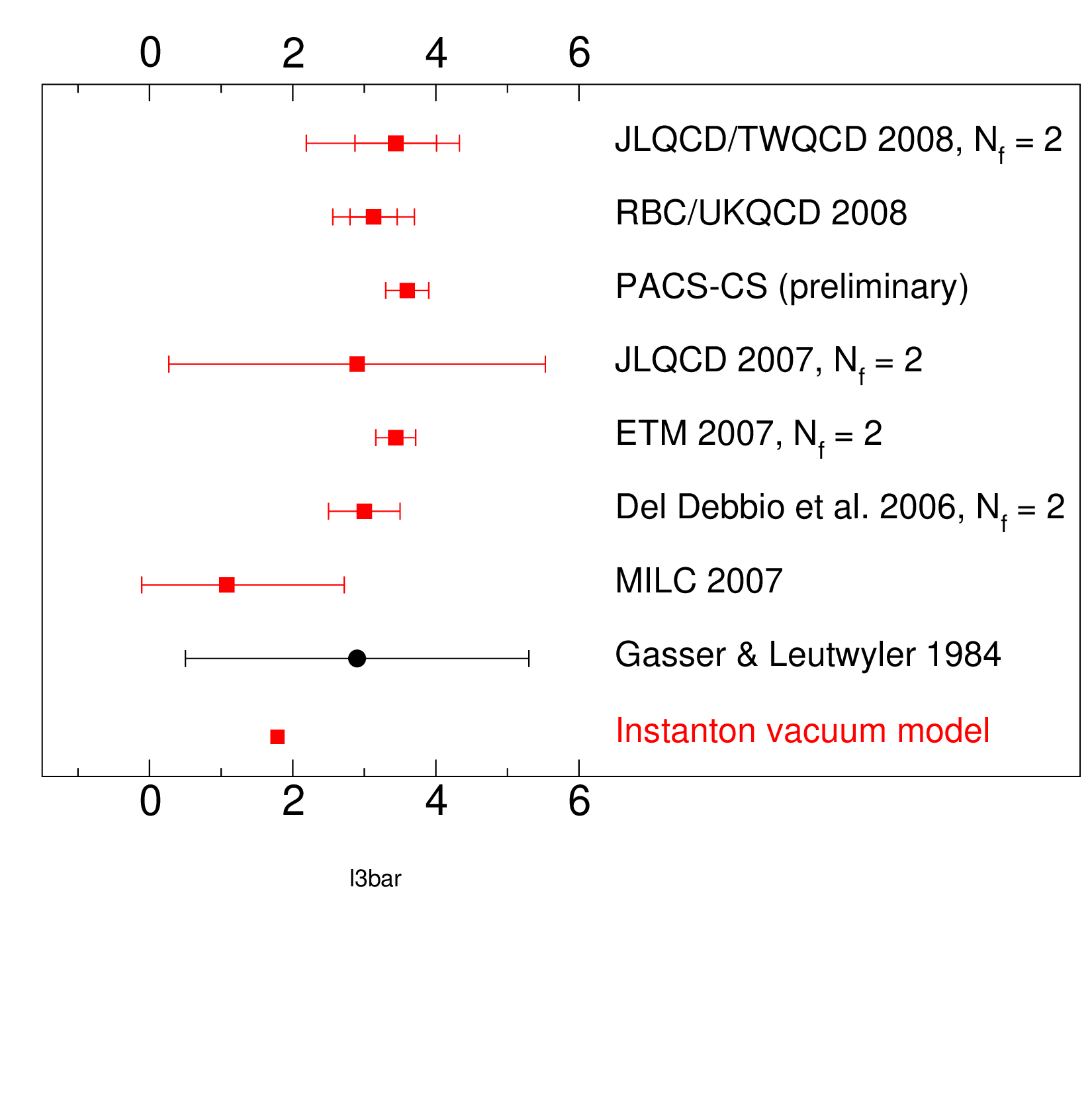}
\includegraphics[scale=0.3,angle=-90]{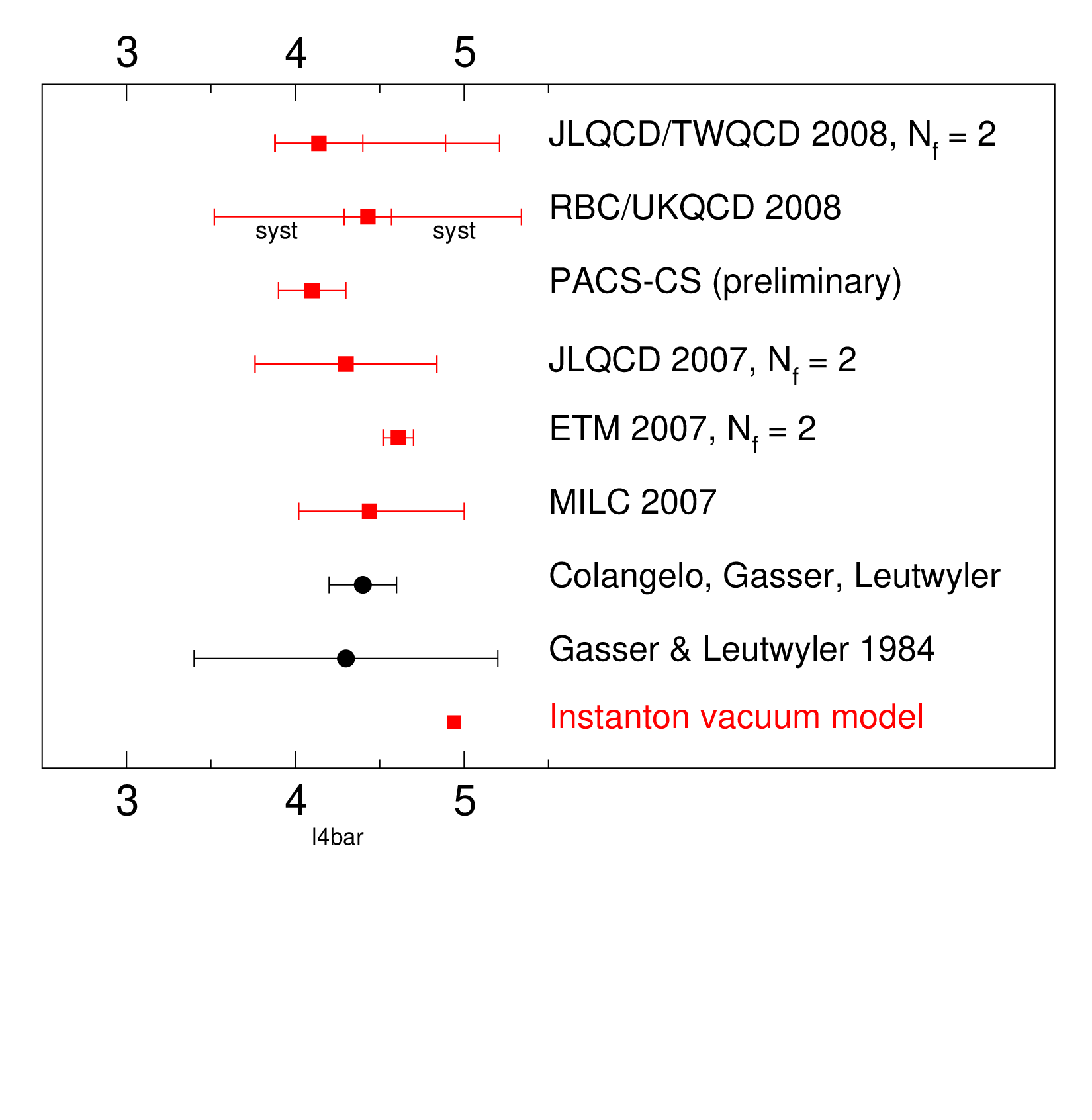}}
\vskip -1.5cm

{\bf Conclusion and outlook.} We established a reliable theoretical framework for the evaluation the $\chi$PT low-energy constants, which provide the understanding of pion physics in QCD.
The calculated constants $\bar l_3$, $\bar l_4$ are in reasonable agreement with lattice results and phenomenological  estimates. 
The calculations of all other constants and the extension to the $N_f=3$ case are on the way.

{\bf Acknowledgements.} The results was obtained in the collaboration with K.~Goeke and M.~Siddikov~\cite{Goeke:2007bj}.
We would like to thank P. Pobylitsa for seminal discussions and P. Bowman for providing us with the lattice data on $M(m)$-dependence.
 The work has been partially supported by the DFG-Graduiertenkolleg Dortmund-Bochum, by the Verbundforschung of BMBF and by the bilateral Funds DFG-436 USB 113/11/0-1 between Germany and Uzbekistan.

\end{document}